\documentclass[]{iopart}

\usepackage{graphicx}
\usepackage{ulem} 	
\usepackage{booktabs}	
\usepackage{upgreek}
\usepackage{subfig}
\usepackage{slashbox}
\usepackage{multirow}
\begin{document}
\title[Suspension platform interferometer for the AEI 10\,m
prototype]{Suspension platform interferometer for the AEI 10\,m prototype:
concept, design and optical layout}

\author{K~Dahl$^1$, O~Kranz$^1$, G~Heinzel$^1$, B~Willke$^1$, K~A~Strain$^{1,2}$,  S~Go\ss ler$^1$
 and K~Danzmann$^{1}$}

\address{$^1$ Max-Planck-Institut f\"ur Gravitationsphysik
(Albert-Einstein-Institut) and Institut f\"ur Gravitationsphysik, Leibniz
Universit\"at Hannover, 30167 Hannover, Germany\\
$^2$ SUPA School of Physics $\&$ Astronomy, University of Glasgow, Glasgow, G12
8QQ, United Kingdom}

\ead{katrin.dahl@aei.mpg.de}

\begin{abstract}

At present a 10\,m prototype interferometer facility is being set up at the AEI
Hannover. One unique feature of the prototype will be the suspension platform
interferometer (SPI). The purpose of the SPI is to monitor and stabilise the
relative motion between three seismically isolated optical tables. The in-vacuum
tables are suspended in an L-shaped configuration with an arm length of
11.65\,m. The design goal of the SPI is to stabilise longitudinal differential
displacements to a level of 100\,pm/$\sqrt{\mathrm{Hz}}$ between 10\,mHz and
100\,Hz and relative angular noise of 10\,nrad/$\sqrt{\mathrm{Hz}}$ in the same
frequency band. This paper covers the design aspects of the SPI,
e.g.~cross-coupling between the different degrees of freedom and fibre pointing
noise are investigated. A simulation is presented which shows that with the chosen 
optical design of the SPI all degrees of table motion can be sensed in a 
fully decoupled way. Furthermore, a proof of principle test of the SPI sensing scheme
is shown. \end{abstract}

\submitto{\CQG}

\section{Introduction}

The AEI 10\,m prototype interferometer~\cite{Goss} aims not only at testing and
developing new techniques for future upgrades of gravitational-wave detectors,
it is also a testbed for ultra-low displacement noise experiments such as to
measure the standard quantum limit of interferometry or performing tests for the
GRACE follow-on experiment~\cite{Dehn}. The prototype has an L-shaped vacuum
envelope with an arm length of about 10\,m. At each end and in the corner of the
`L' a tank is located, housing a 1.75\,m~$\times$~1.75\,m optical table
supported by a seismic attenuation system derived from the HAM-SAS
table~\cite{Stoc}. Although these tables have been designed to provide excellent
seismic isolation, there will still be a certain amount of residual motion
between the three tables. The task of the suspension platform interferometer
(SPI) is to sense this relative motion and to allow suppression by actuators which 
are part of the seismic attenuation system of each optical table.

One of the planned experiments is to perform tests for a GRACE follow-on
mission. GRACE~\cite{Tapl} consists of two satellites, its goal is to map the
Earth's gravity field. The orbital motion of each satellite senses local
variations of the Earth's gravity field. This leads to small fluctuations of the
inter-satellite distance. In order to test laser interferometers for GRACE
follow-on missions, it is planned to perform experiments within the AEI 10\,m
prototype where the suspended tables represent satellites. By using signals of
the SPI, the tables will be moved on purpose by a well-known amount. This
experiment sets the design goal for the SPI in the lower frequency range, which
is less than 100\,pm/$\sqrt{\mathrm{Hz}}$ between 10\,mHz and 100\,Hz for
longitudinal table motion and 10\,nrad/$\sqrt{\mathrm{Hz}}$ for angular noise in
the same frequency band.

The main experiment that will be set up on the tables of the AEI 10\,m prototype
facility is a Michelson interferometer with Fabry-Perot arm cavities. The
scope of this interferometer is to probe at and beyond the standard quantum
limit of interferometry~\cite{Cave}. Therefore, all classical noise sources,
e.g.~seismic noise, have to be sufficiently reduced to leave the sensitivity
exclusively limited by quantum noise in the measurement band of interest. A
straightforward way to reduce seismic noise is to suspend all optics of the
interferometer. The SPI and the table position control act to reduce the relative motion of the tables, and
hence stabilises the suspension positions of the suspended interferometer
components that are supported by the tables. This stabilisation decreases the 
required force and hence the noise of the voice coil actuators that are an integral 
part of the position control of each suspension. The reduced relative velocity of 
the components also makes it easier for the interferometer control system to acquire lock.

\section{Introduction to suspension platform interferometry}

The idea to set up an ancillary interferometer beside the main interferometer to
improve lock acquisition and the operation of the main interferometer,
especially at low frequencies, was proposed by Drever~\cite{Drev}. He suggested
to install and lock an ancillary interferometer at the intermediate stage of the
suspended mirrors of the main interferometer to reduce their residual rms
motions. This idea was tested by Aso et al~\cite{Aso}. They measured 40\,dB
noise reduction between 0.1\,Hz and 1\,Hz. In an experiment conducted by Numata
and Camp~\cite{Numa}, the relative longitudinal and yaw motions between two
hexapods separated by 1\,m were measured by use of three homodyne Michelson
interferometers. The longitudinal displacement was stabilised to
1\,nm/$\sqrt{\mathrm{Hz}}$ at 1\,mHz. 

The SPI described in this paper is unique in the sense that
\begin{itemize}
  \item the SPI will monitor the relative motion between three
suspended optical tables (weight of about 1\,t each and separated by about
10\,m). 
 \item the SPI will provide continuous error signals over a wide
($\gg$$\lambda$) operating range to track the relative table motions.
 \item the SPI will sense all degrees of freedom except for roll around the beam
axis.
\end{itemize}

Several optical configurations can be used to sense relative motions between two
objects, e.g.~a Fabry-Perot cavity or an optical configuration using a
pseudo-random noise code~\cite{Shad,Devi}. The advantage of a Fabry-Perot
cavity is that it is a fairly simple setup. The main disadvantage and the reason
why it was not an option for the SPI is the limited sensing range of one cavity 
linewidth. We decided to use heterodyne Mach-Zehnder interferometry to
monitor the relative motions of the three suspended tables. This provides the
desired constant sensing performance at any table position over many optical
wavelengths. Furthermore, we are able to benefit from in-house experience gained
with the experiments for LISA Pathfinder~\cite{Nama}.

The optical layout of the SPI is shown in figure~\ref{fig:all}. It is the result
of simulation work taking into consideration several requirements. In this paper
the layout is presented first, followed by details of the simulation from which
it resulted. 

The following nomenclature is used in this paper: all photodiodes are
abbreviated with PD followed by another letter that indicates to which
interferometer the photodiode belongs (D for diagnostic, R for reference, S for
south, and W for west). All beam splitters are abbreviated with BS, all beam
recombiners with BR, and all mirrors with M. All optics are placed on the
central table except for mirror MS which is placed on the south table and mirror MW which is
placed on the west table. The reference beam "beamR" is confined to the
baseplate on the central table, while parts of the measurement beam "beamM" also
travel to the south and west tables.

\section{Optical layout}

\begin{figure}[t]
 \centering
 \includegraphics[width=.99\textwidth]{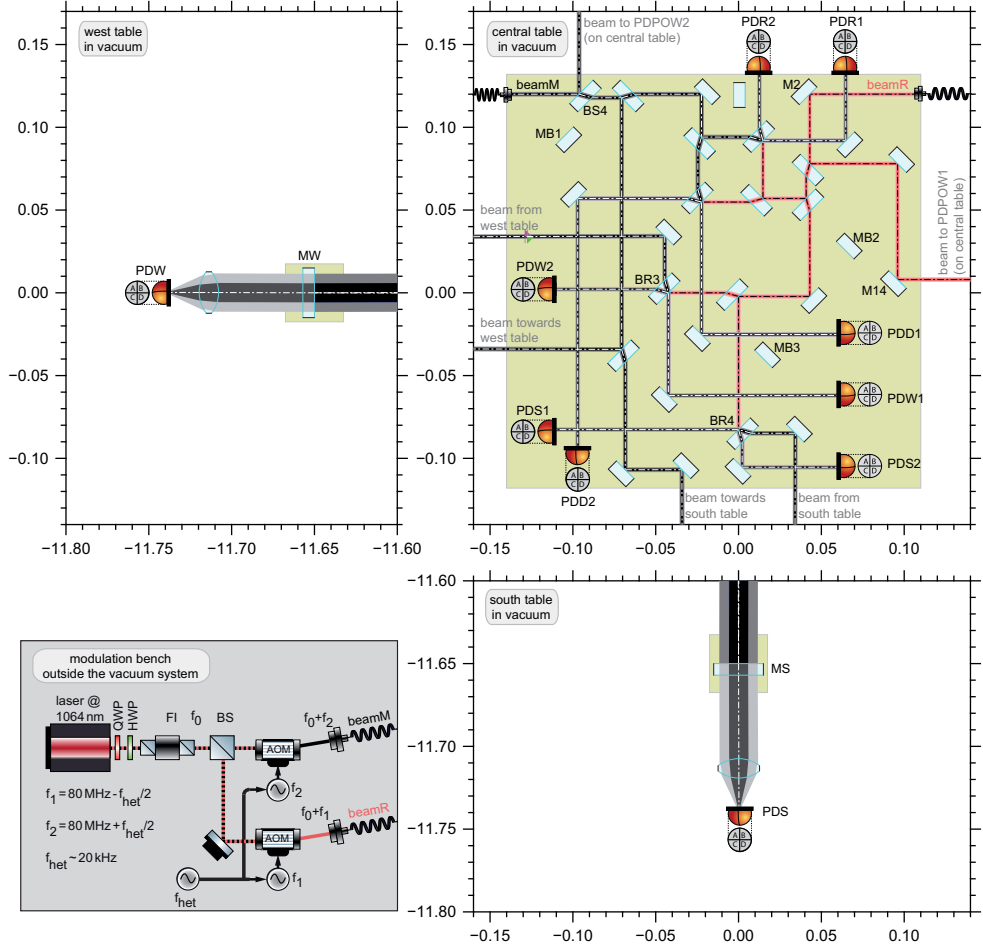}
 \caption{The optical layout of the SPI. The modulation bench which is placed
outside the vacuum system is depicted in the lower left. On the modulation bench
the laser light is prepared and coupled into optical fibres. These fibres
deliver the beams to the measurement bench which is placed on the central table
in the vacuum system (see upper right box). The reference beam, beamR, is drawn
in red; the measurement beam, beamM, in black. All scales show the distance from 
the centre of the central table in meter. x~=~0\,m, y~=~0\,m is the centre of the 
central table; x~=~-11.65\,m, y~=~0\,m is the centre of the west table (upper left box), 
and x~=~0\,m, y~=~-11.65\,m is the centre of the south table (lower right box).}
 \label{fig:all}
\end{figure}

The SPI consists in total of four non-polarising heterodyne Mach-Zehnder
interferometers. All interferometers share the same optical path on the
modulation bench, which is shown in the lower left box of figure~\ref{fig:all}.
Here, the laser beam is prepared and coupled into optical fibres. The fibre
outcouplers are mounted on the measurement bench located on the central table inside the vacuum system. In
contrast to the modulation bench which uses conventional optical mounts, the
measurement bench is quasi-monolithic and thus much more stable in terms of
mechanical and thermal drifts. All displacement measurements are performed on
the measurement bench. 

\subsection{Modulation bench}

The modulation bench housing the laser is located outside of the vacuum system.
We choose to work with a continuous-wave laser of a wavelength of 1064\,nm and
aim to reduce the SPI sensing noise due to laser frequency fluctuations to
10\,pm/$\sqrt{\mathrm{Hz}}$ at 10\,mHz, to make sure that laser frequency noise
will not limit the SPI sensitivity. Since the arm length mismatch of two of the
four Mach-Zehnder interferometers is about 23\,m, the laser frequency stability
has to be better than 120\,Hz/$\sqrt{\mathrm{Hz}}$ at 10\,mHz. We choose a
commercially available iodine-stabilised Nd:YAG laser~\cite{Inno}. 

The laser light is split into two paths (at beam splitter BS in the lower left
box of figure~\ref{fig:all}). After that, each of the two beams is
frequency-shifted by an acousto-optic modulator (AOM) operating near 80\,MHz,
with a frequency difference between the two channels set to the desired
heterodyne frequency, which has been chosen to be around 20\,kHz in order to
provide a control bandwidth up to 100\,Hz. The light (beamR and beamM in
figure~\ref{fig:all}) is coupled into two 20\,m long polarisation-maintaining
single-mode optical fibres. These are fed into the vacuum system, and beamR and
beamM are delivered to fibre couplers mounted on the measurement bench.

\subsection{Measurement bench}

The measurement bench is located on the central table (inside the vacuum system)
and holds the mechanically and thermally ultra-stable part of the four
interferometers. To ensure that thermal drifts are kept sufficiently small, the
measurement bench is made of Clearceram\textsuperscript{\textregistered}{}-Z HS,
an ultra-low expansion material with a coefficient of thermal expansion of
(0.0$\pm$0.2$)\cdot$10$^{-7}$\,K$^{-1}$ and a zero-crossing at room
temperature~\cite{Naka}. All beam splitters and mirrors except MS and MW are
hydroxy-catalysis bonded~\cite{Elli} to the 250\,mm$\times$250\,mm surface of
the 30\,mm thick measurement bench. The remote mirrors MS and MW are bonded on
two small cuboids of Clearceram\textsuperscript{\textregistered}{}-Z HS of
35\,mm$\times$35\,mm and 30\,mm height. MS is placed on the south table, MW is
placed on the west table.

Except for MS and MW, all mirrors and beam splitters are flat. In order to
achieve a high interferometric contrast and a good sensitivity of the
differential wave-front sensing signal (DWS) (for further information on DWS see
section~\ref{sec:processing}), the radius of curvature of MS and MW has been
chosen such that the waist of the two beams is at the recombining beam splitters
(BR1, BR2, BR3, BR4). Thus, the beam radius and curvature of the beam's wave
front is the same for both interfering beams. This leads to a high
interferometric contrast. MS and MW have a radius of curvature of -11.8\,m,
i.e.~they are concave. 

\begin{figure}[t]
 \centering
 \includegraphics[width=.99\textwidth]{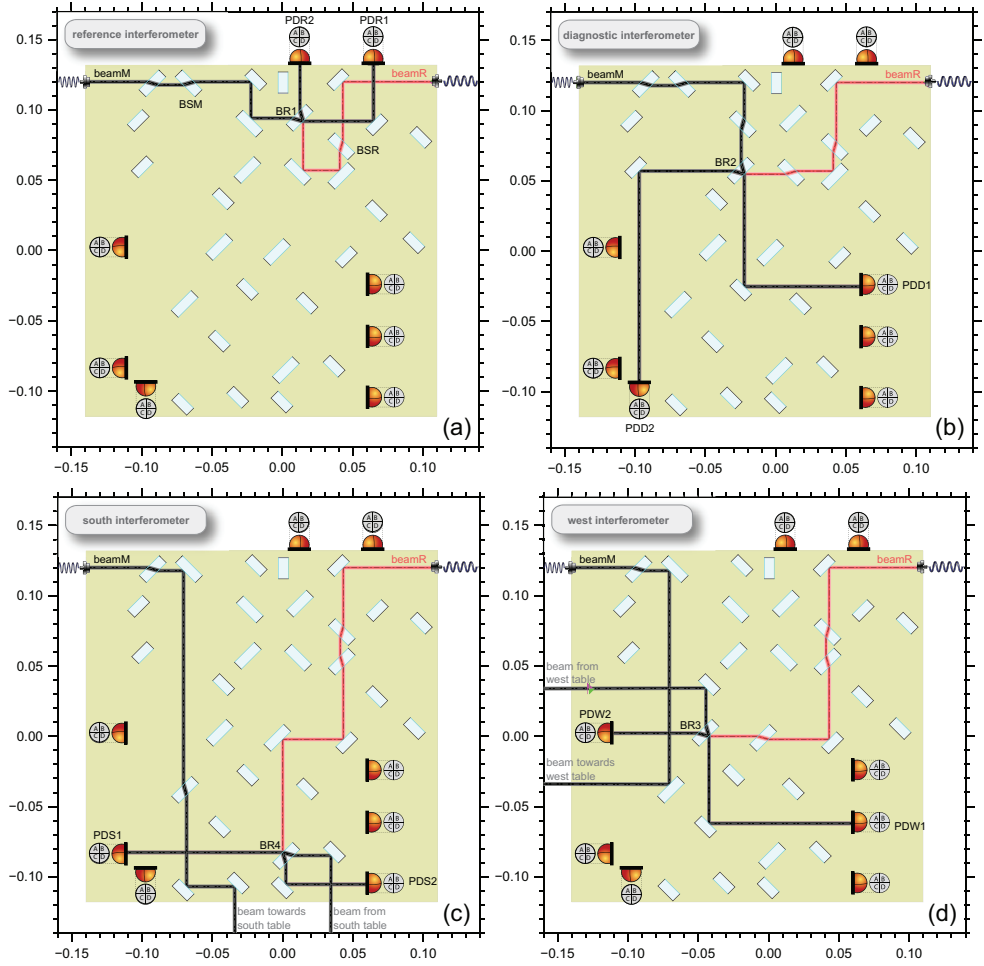}
 \caption{Each of the figures shows the optical path of one of the four interferometers 
on the measurement bench.}
 \label{fig:4ifo}
\end{figure}

One of the four interferometers is the reference interferometer (see
figure~\ref{fig:4ifo}(a)). It is used to cancel common mode fluctuations by
subtracting them from all other interferometer outputs. The reference
interferometer measures all path length equivalent fluctuations that have been
picked up between the first beam splitter BS on the modulation bench and the
recombing beam splitter BR1 on the measurement bench. Therefore, beamR and beamM
are recombined at BR1 and their beat note is detected at photodiodes PDR1 and
PDR2. Since  all interferometers share the same optical paths between the modulation 
bench and beam splitter BSM and BSR on the measurement bench, all interferometers 
are affected by the same phase noise. Path length equivalent fluctuations are caused by, 
e.g.~drifts of the conventional mirror mounts on the modulation bench and by stress on 
and movements of the optical fibres. The optical fibres are flexible but need to be installed 
loosely so that they do not compromise the excellent seismic isolation of the optical
tables. 

The second interferometer, the diagnostic interferometer, is also entirely
confined to the 250\,mm$\times$250\,mm baseplate (see figure~\ref{fig:4ifo}(b)).
This diagnostic interferometer is the most striking difference of the
measurement bench presented in this paper compared to the measurement bench
previously shown in~\cite{Dahl}. The diagnostic interferometer in conjunction
with the reference interferometer will be used for debugging purposes and to
determine the sensitivity of the bonded setup since the reference and diagnostic 
interferomter nominally have the same signals. The diagnostic and reference 
interferometers are designed such that their optical arm length difference is zero. 
Within a Mach-Zehnder interferometer of equal arm length, laser frequency 
noise does not couple into the interferometric output signal.  In this way, it 
is possible to determine the limit to which the south and west tables can be 
stabilised relative to the central table. A displacement noise measurement 
between the reference and diagnostic interferometers gives the sensor noise 
(except laser frequency noise) of the whole SPI. Hence, the limit for stabilising 
the south and west tables to the central table for an ideal feedback loop. For the 
diagnostic interferometer the light of beamR and beamM is recombined at BR2 
and detected at photodiodes PDD1 and PDD2. 

The other two interferometers, namely the west and south interferometers (see
figure~\ref{fig:4ifo}(c)-(d)), measure the displacement and angular deviation of 
the south/west table relative to the central table. These two interferometers are 
the so-called measurement interferometers. They are of unequal arm length. One 
arm of the interferometers (the arm carrying reference beamR) is entirely on the 
central baseplate. The other arm which is carrying beamM is leaving the central
baseplate towards the west table in case of the west interferometer and to the
south table in case of the south interferometer. BeamM is reflected back to the
measurement bench by MS and MW, respectively. 

BeamR and beamM recombine at BR3 and BR4 for the west and south interferometers,
respectively. The interference pattern of the west interferometer is detected by
photodiodes PDW1 and PDW2, the beat note of the south interferometer at PDS1 and
PDS2. A motion of the south/west table can be monitored by the south/west
interferometer since the length of the interferometer arm carrying beamM is
changing while the south/west table and/or central table is moving, whereas the
other arm carrying beamR is of constant length.

The beams reflected by beam splitter BS4 and mirror M14 (see
figure~\ref{fig:all}) are used for power stabilisation and to monitor pointing
noise of the fibre injectors. For further details on pointing noise of fibre
injectors see subsection~\ref{sec:pointing}. The light being transmitted by MS
and MW due to the residual transmission of the dielectric coating is detected
by quadrant photodiodes. These are operated at DC and are used to get
additional information of the south and west tables' motion relative to the
central table. 

Substrates MB1, MB2, and MB3 (see figure~\ref{fig:all}) are not part of any of
the four interferometers. They are needed as reference within the manufacturing
process of the quasi-monolithic measurement bench. The alignment of some of the
to-be-bonded mirrors and beam splitters is done either by use of a brass
template or by use of a coordinate measurement machine. Between the bonding of
the several optics the brass template has to be removed. MB1, MB2, and MB3
define reference points to which the template is realigned when it is placed
again onto the measurement bench. For more details on the bonding process and
the accuracy achieved see~\cite{Dahl_bond}.

\section{Signal processing}
\label{sec:processing}

The SPI should not only be able to monitor longitudinal relative table motion,
but also pitch and yaw motion of the tables. Hence, all photodiodes used for the
SPI are quadrant photodiodes. To save as much space as possible for other
experiments that will be conducted within the AEI 10\,m prototype, the
photodiodes are mounted on the Clearceram\textsuperscript{\textregistered}{}-Z
HS  baseplate. While the photodiodes themselves are inside the vacuum system,
the related electronics for signal processing is outside. The signals produced
by the recombined beams on photodiodes PDR1, PDR2, PDD1, PDD2, PDS1, PDS2, PDW1,
and PDW2 are routed to a phasemeter as developed for LISA Pathfinder
experiments~\cite{Hein}. The photocurrents from the other diodes, i.e.~PDPOW1,
PDPOW2, PDS, and PDW, are routed to signal conditioning electronics.

Each channel of signal conditioning electronics is basically a transimpedance
amplifier that converts photo currents into voltages. As well as the voltage for
each quadrant, outputs are provided for the sum of all quadrants, the difference
between upper and lower quadrants, and the difference between left and right
quadrants. All outputs are fed from the signal conditioning electronics into a
realtime Control and Data System (CDS) which was developed for the LIGO
project~\cite{Shoe} and adopted for the AEI 10\,m prototype. Within the CDS all
signals are digitised and used to monitor and control beam pointing and the
tables' relative motion.

The signals from the photodiodes (listed above) which detect recombined beams,
are fed to the phasemeter where they are converted to voltages, digitised and
Fourier-transformed at the heterodyne frequency by a single-bin discrete Fourier
transform. The phasemeter output values for each photodiode quadrant are the DC 
signal, and the real and imaginary part of the complex amplitude of the photodiode 
signal at the heterodyne frequency. The argument of the complex amplitude is the 
phase signal. The phasemeter outputs are transmitted to the CDS via a microcontroller-based
phasemeter interface. Within the CDS the signals are combined such that for each
quadrant photodiode longitudinal phase information, differential wave-front
sensing (DWS)~\cite{Morr} signals for pitch and yaw, differences of left and
right as well as bottom and top of the DC photodiode signals, and the
interferometric contrast are available. These signals are used to derive error
signals for feedback control of the optical tables via voice-coil actuators as
described below.

\section{Simulation work for the design of the optical layout}

Within the process of designing, we minimised coupling of table rotation 
to the translational motion and investigated the coupling factors. Further 
simulations were carried out to determine whether the beam pointing noise 
of the fibre injectors might spoil the SPI signals. All simulations including 
the design of the optical layout have been done with Optocad~\cite{opto} 
and partly with Ifocad. 

\begin{figure}[t]
  \centering
  \subfloat[Explanation of table central axis and connecting line.]{ 
    \includegraphics[width=.35\textwidth]{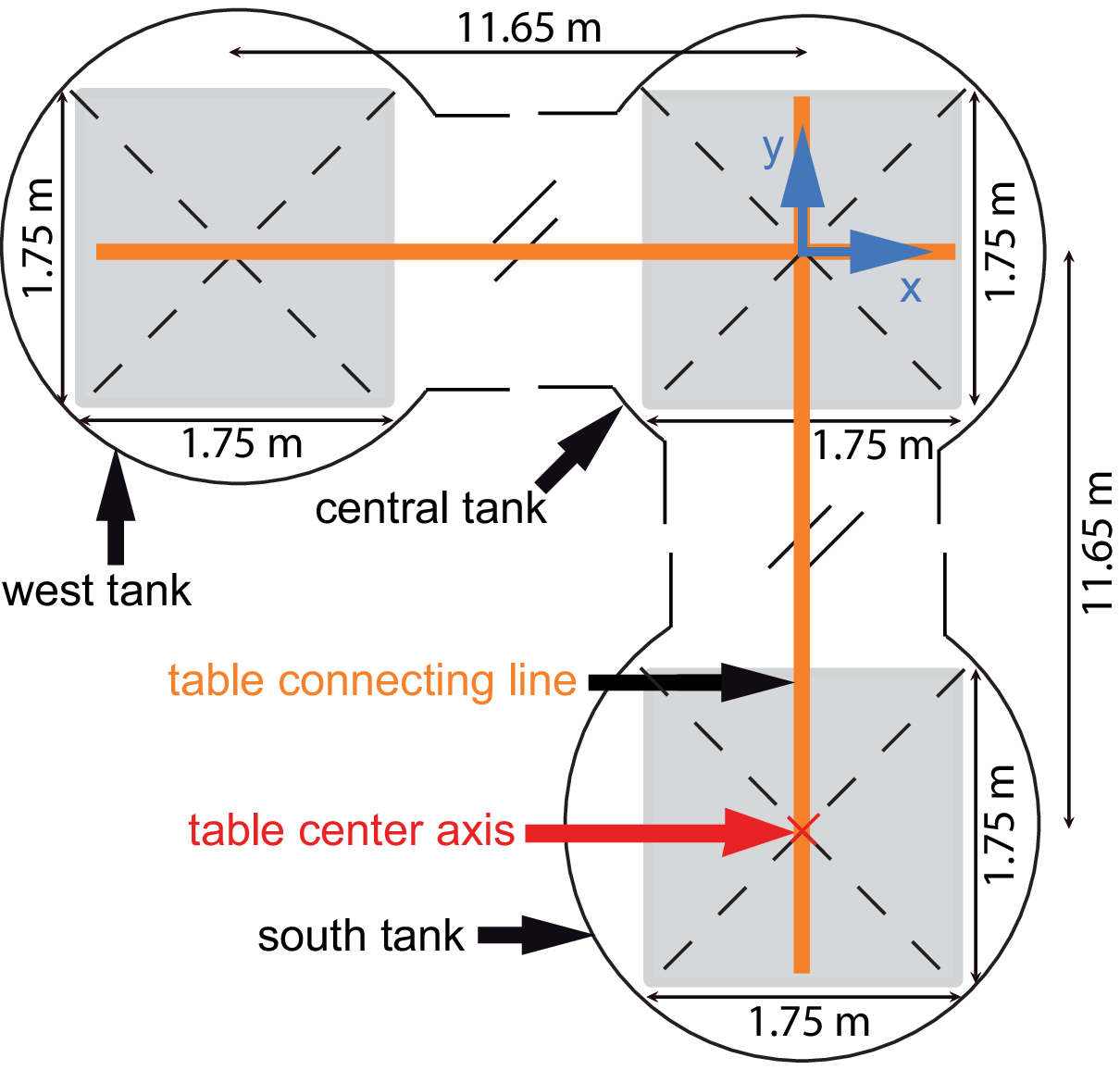}
\label{fig:table_connecting_line}
  }
  \subfloat[Coupling of pointing noise of beamM to the output signals.]{
    \includegraphics[width=.65\textwidth]{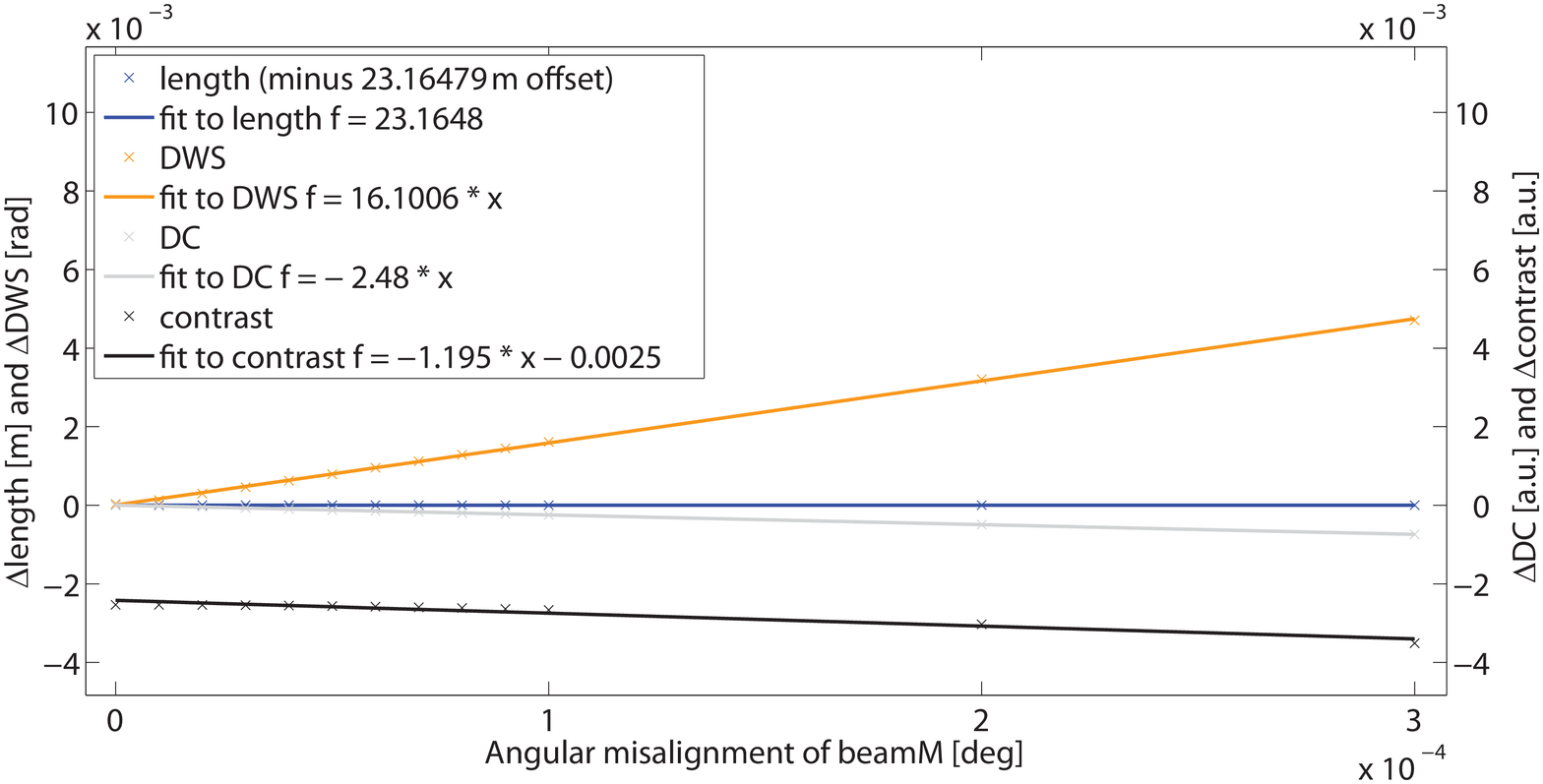}
\label{fig:pointing_noise_pds1-pdr2}
      }
  \caption{(a) This sketch explains what is meant by table central axis and
connecting line. The coordinate system used throughout all simulations is given
in blue. (b) This graph shows simulation results how pathlength, DWS, DC, and 
contrast change in dependence of angular misalignment of the fiber injector of 
beamM. The signals of photodiodes PDS1 and PDR2 have been subtracted and 
a fit to the data has been done.}
  \label{fig:X}
\end{figure}

Due to the symmetric table design the table rotates around its central axis. 
Thus, coupling of table rotation to the translational motion is minimised if 
the far mirrors MS and MW (see figure~\ref{fig:all}) are placed in the centre 
of the south and west tables, respectively, and the recombining beam splitters 
BR3 and BR4 are placed along the table centre connecting line (see figure~\ref{fig:table_connecting_line}).

To make the control of the west and south interferometers as close to identical 
as possible, both the west and south interferometers, have a design value of arm 
length mismatch of 23165\,mm. The arm length mismatch of the reference and
diagnostic interferometers is designed to be identical, too. The nominal arm length 
mismatch of the  interferometers is in reality limited by the accuracy of the assembly
process of the measurement bench, i.e.~in the order of a few $\upmu$m.

\subsection{Coupling factors}
\label{sec:sensing}

To check how well the different relative table motions are decoupled we
calculated the coupling factors for the west and south interferometers. A 
coupling factor is the first derivative of a signal by a degree of freedom (DoF). 
A set of coupling factors calculated for photodiode PDS2 of the south 
interferometer is shown in table~\ref{tab:sens_matrix_design_PDS2_fit}.

To obtain the coupling factors shown in this table, mirror MS was moved in 50 steps
$\pm$100\,$\upmu$m or $\pm$4\,$\upmu$rad away from the well-aligned
configuration. This movement of MS, which is in reality a motion of the south
table, was performed in two translational degrees of freedom and in the yaw
degree of freedom since Optocad can cope with two dimensions only. At each
single position of MS pathlength, DWS, DC, and interferometric contrast of PDS2 were
calculated. The coupling factors were then found by taking
the ratios of all of the output signals to each of the input motions.

The length channel is the dedicated channel to monitor relative longitudinal
motions of the tables. From the column called 'Length' in
table~\ref{tab:sens_matrix_design_PDS2_fit} one can clearly see that almost no
coupling of the transversal and rotational motion of the south table into the
length signal occurs. Thus, it will be possible to distinguish longitudinal
relative table motion from any other possible table motion. In the experiment
the longitudinal degree of freedom will be the first one to be controlled. A rotation of the south table can be easily monitored by the DWS
signal since the factor for DWS rotation is an order of magnitude larger for
rotation than for any other possible table motion (see third column in
table~\ref{tab:sens_matrix_design_PDS2_fit}). Rotation will be the second degree
of freedom that will be stabilised. The transversal table motion will be
monitored by the DC channels. Though the factor is much larger for rotation
than for transversal motion, it is still possible to distinguish between the two
motions because the rotation of a table was stabilised before by using the DWS
signals. Thus, the only remaining motion that can produce a signal in the DC 
channel is the table's transversal motion.

It will be easily possible to distinguish between rotational and longitudinal
relative table motion. Thus, the optical layout as depicted in
figure~\ref{fig:all} is suitable to achieve the SPI's design goal. During the
installation of the SPI the coupling factors will be determined experimentally by
moving the table and measuring the table's position by LVDTs (linear variable
differential transducer).

\begin{table}
\caption{\label{tab:sens_matrix_design_PDS2_fit} Coupling factors between motion in the table degree of freedoms and SPI sensing 
signals. These coupling factors were calculated for simulated motions of mirror MS
and detection at photodiode PDS2. A motion of MS in the simulation is equivalent
to motion of the south table in the experiment. The values for length, DWS, and
DC are slopes with a zero-crossing at the properly aligned configuration for all
investigated degrees of freedom.}
\begin{indented}
\item[]
\begin{tabular}{@{}lrrr}

\br
\multicolumn{1}{r}{Signal}  &Length & DWS & DC    \\
&\crule{3}\\
DoF& (m/m) & (rad/m)  & (a.u./m)   \\ 
\mr
Transversal & 3.5e-09 & -3.2e+02 & 2.3e+03 \\ 
Longitudinal & 2.0e+00 & 1.8e+00 & -8.9e-02 \\ 
&\crule{3}\\
 &  (m/rad)  & (rad/rad) & (1/rad)   \\ 
&\crule{3}\\
Rotational & 1.2e-07 & -3.7e+03 & 2.8e+04   \\ 

\br
\end{tabular}
\end{indented}
\end{table}

\subsection{Pointing noise of fibre injectors}
\label{sec:pointing}

A potential noise source is the pointing noise of the fibre injectors. We use a custom-made 
vacuum compatible version of the adjustable optics holder AAH 5 axes by miCos.
They are
the only potentially unstable components of the in-vacuum setup since all
mirrors and beam splitters are bonded onto an ultra-low expansion baseplate. The
idea of having the reference interferometer is to measure all potential
instabilities and subtract them from the measurement interferometers, which are
the west and south interferometers. In the measurement interferometers the
optical path length of beamM is about a factor of 80 longer than in the
reference interferometer, i.e.~the measurement interferometers are much more
sensitive to pointing noise of beamM than the reference interferometer. Thus,
pointing noise of a fibre injector could misleadingly be understood as a
relative table motion. 

In the simulation beamR was fixed and beamM was moved in yaw. The signals of the
photodiodes of the south, west, and reference interferometers were then
evaluated. Figure~\ref{fig:pointing_noise_pds1-pdr2} shows exemplarily for PDS1
and PDR1, that beam pointing noise couples into the DWS signal and cannot be
removed by subtracting the signal of the reference interferometer from the
signals of the south interferometer. To substract the signals of the south
interferometer from the signals of the west interferometer is not an option
since then no monitoring of the relative table motion would be possible
anymore. 

The design goal for angular noise in the SPI sensing is 10\,nrad/$\sqrt{\mathrm{Hz}}$ 
from 10\,mHz up to 100\,Hz. To check up to which value of fibre injector pointing 
noise this can be achieved, a linear fit to the DWS output values as depicted in
figure~\ref{fig:pointing_noise_pds1-pdr2} was performed. According to the fit to
the DWS signal the beam pointing noise has to be below
11\,prad/$\sqrt{\mathrm{Hz}}$ to reach the design goal for angular
noise.

Thus, the pointing noise of the fibre injectors will be monitored by additional
non-coherent quadrant photodiodes PDPOW1 and PDPOW2 (see grey labels in upper
right box of figure~\ref{fig:all}) and probably corrected in signal
post-processing. The optical pathlength to PDPOW1 and PDPOW2 will be about
70\,cm. The pathlength from the beam injector to PDR1 and PDR2 is only about 30
cm. This results in a 7/3 better sensitivity for the DC signal in beam pointing for 
PDPOW1 and PDPOW2 than for PDR1 and PDR2.

The fit to the phase signal in figure~\ref{fig:pointing_noise_pds1-pdr2}
indicates that there is no coupling from pointing noise of beamM into the
longitudinal phase channel. However, Guzm\'an Cervantes et al.~\cite{Guzm} have
measured that angular noise couples to some extent into the longitudinal phase
signal. They further demonstrated that by determining the coupling factors, the
angular noise can be well subtracted from the longitudinal phase channel. 

If beam pointing limits the sensitivity of the SPI, we will post-process the
data and the current fibre injectors will be replaced by monolithic fibre
injectors. These injectors were not available at the beginning of the SPI
construction. Dedicated space to bond the monolithic fibre injectors on the
measurement bench has been left in front of beam splitter BS4 and mirror M2.

\section{Test setup}

\begin{figure}[t]
  \centering
  \subfloat[longitudinal shift of MS]{ 
    \includegraphics[width=.5\textwidth]{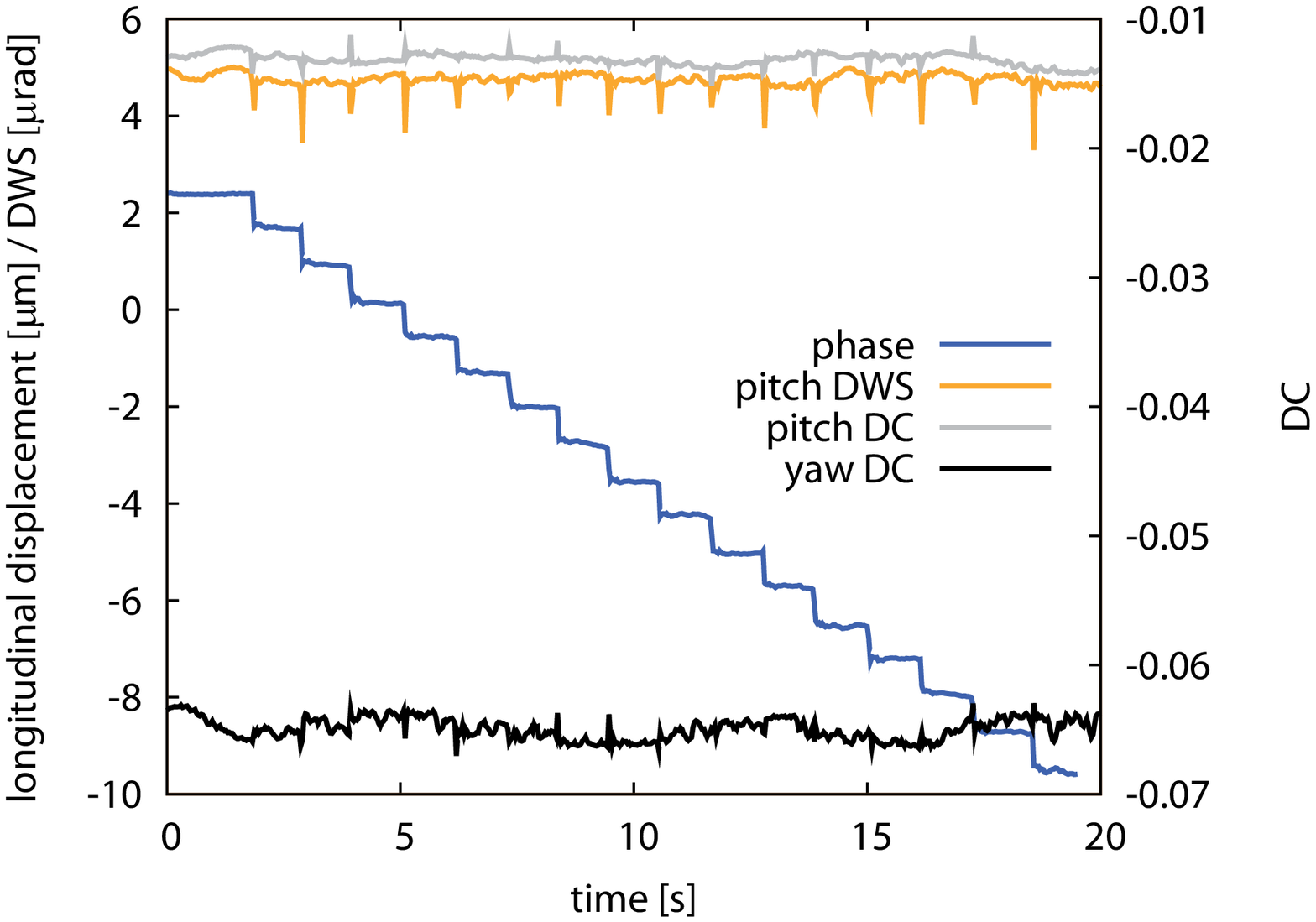} \label{fig:MS_southwards}
  }
  \subfloat[MS changes in pitch]{
    \includegraphics[width=.5\textwidth]{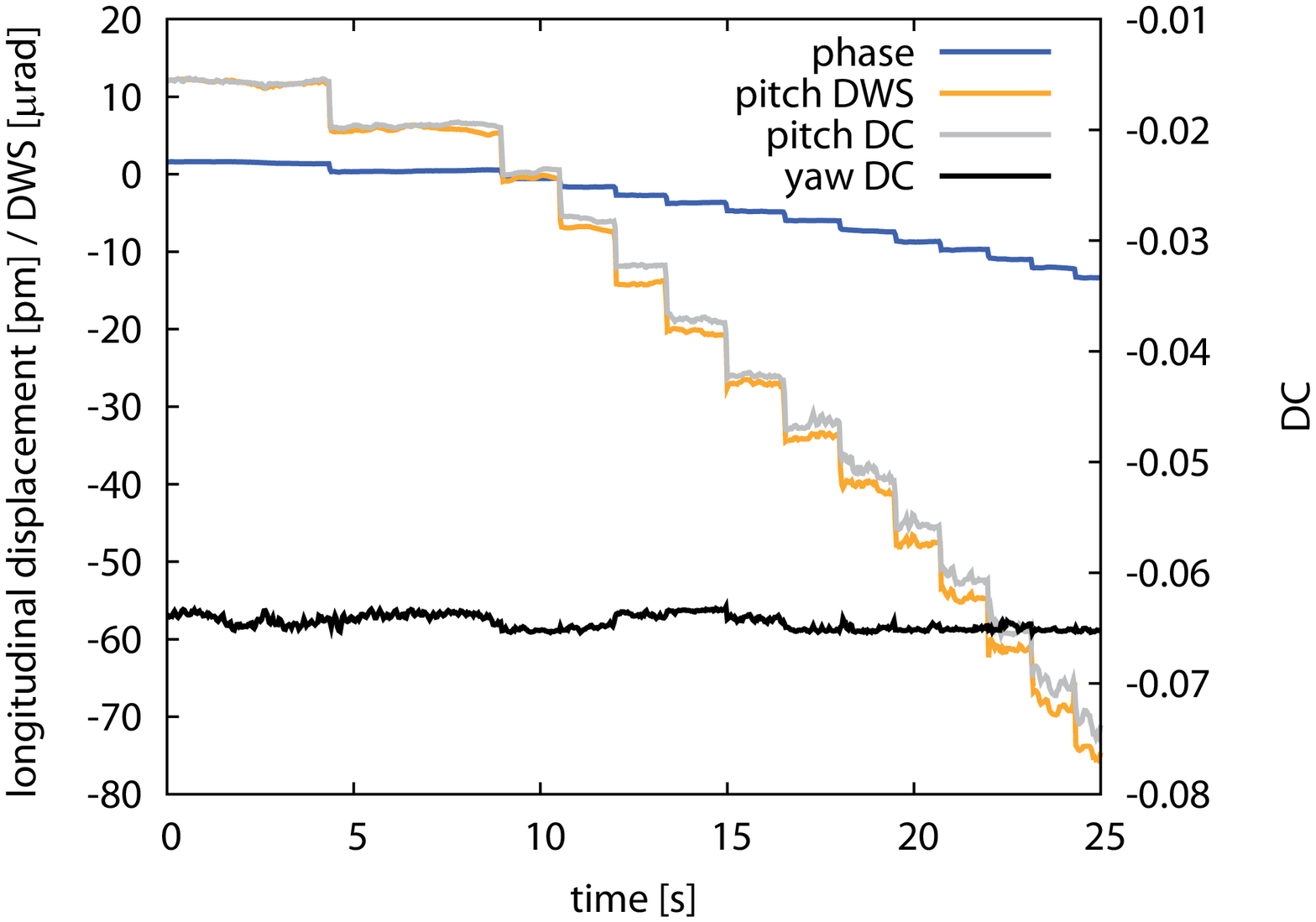} \label{fig:MS_pitch}
  }
  \caption{Signals of the test setup (a) if mirror MS is shifted away from the
residual optical setup and (b) if mirror MS is moved in the pitch degree of freedom.}
  \label{fig:test_setup}
\end{figure}

During the design process we have built a test setup to measure the relaitve motion 
between two standard optical tables. The test setup includes reference and south 
interferometers as well as an additional Michelson interferometer for calibration. All 
optics other than mirror MS belonging to the south interferometer were on one optical 
table. Only MS was on another optical table. The arms of the reference interferometer
were of about equal length. The arm length difference of the south
interferometer was about 3\,m. The radius of curvature of mirror MS was -2\,m.
MS was mounted in a piezo-driven mirror mount and placed on two linear stages.
In this way, MS could be moved in pitch and yaw as well as longitudinal and
transversal relative to the optical table hosting the rest of the south
interferometer. A Michelson interferometer was used to calibrate the
piezo-driven mirror mount and the linear stages. 

Figure~\ref{fig:MS_southwards} shows the signals when MS is moved away from the
rest of the interferometer. There is only a change in the longitudinal signal
and no coupling into the pitch and yaw channels. If MS is moved in pitch (see
figure~\ref{fig:MS_pitch}) the pitch DWS and pitch DC signals clearly indicate a
pitch motion. The small coupling into the longitudinal phase channel is due to
the fact that the beam was not centred on the curved mirror MS.

The results of the test setup confirmed our approach to use a set of heterodyne
Mach-Zehnder interferometers to monitor the relative motion between the 11.65\,m
apart optical tables within the vacuum system of the AEI 10\,m prototype.

\section{Status, conclusion and outlook}
In this paper, we introduced the measurement concept and optical layout of the
SPI for the AEI 10\,m prototype. The results of our test setup show that
technique of heterodyne Mach-Zehnder interferometry is suitable to monitor
relative table motions. Furthermore, simulations regarding the position and
sensitivity of the optical layout of SPI were presented. The results show that
the design of the optical layout is suitable to achieve the design goal of
100\,pm/$\sqrt{\mathrm{Hz}}$ from 10\,mHz up to 100\,Hz for displacement noise
and 10\,nrad/$\sqrt{\mathrm{Hz}}$ in the same frequency band for angular noise. 

Currently, three-quarters of the SPI is already bonded. Around winter 2011/2012, the
SPI is going to be installed in the vacuum system of the AEI 10\,m prototype
after the commissioning of the first two optical tables.

\ack{The authors would like to thank the IMPRS for Gravitational Wave Astronomy,
the AEI LISA group for support, the Excellence Cluster QUEST (Centre for Quantum
Engineering and Space-Time Research) for financial support, and Gudrun Wanner
for providing her simulation code. }


\section*{References}


 \begin{thebibliography}{40}

\bibitem{Goss} Go{\ss}ler S {\it et al} 2010 \CQG {\bf 27} 084023

\bibitem{Dehn} Dehne M, Guzm\'an Cervantes F, Sheard B, Heinzel G and Danzmann K
2009 {\it J.\ Phys.:\ Conf.\ Ser.} {\bf 154} 012023

\bibitem{Stoc} Stochino A {\it et al} 2009 \NIM {\it A} {\bf 598} 737

\bibitem{Tapl} Tapley B, Bettadpur S, Watkins M and Reigber C 2004 {\it Geophys.\ Res.\ Lett.} {\bf 31} L09607

\bibitem{Cave} Caves C M 1980 \PRL {\bf 45} 75--79

\bibitem{Drev} Drever R W P 1987 Outline of a proposed design for a first
receiver for installation in the long-baseline facilities, of Fabry-Perot type
{\it LIGO Document} {\bf T870001-00-R}

 \bibitem{Aso} Aso Y, Ando M, Kawabe K, Otsuka S and Tsubono K 2004 \PL {\it A} {\bf 327} 1

 \bibitem{Numa} Numata K and Camp J 2008 {\it Appl.\ Opt.} {\bf 47} 6832

\bibitem{Shad} Shaddock D A 2007 {\it Opt. Lett.} {\bf 32} 3355

\bibitem{Devi} de Vine G {\it et al} 2009 {\it Opt.\ Express} {\bf 17} 828

 \bibitem{Nama} Mc{N}amara P, Vitale S, Danzmann K and on behalf of the {LISA} 
{P}athfinder {S}cience {W}orking {T}eam 2008 \CQG {\bf 25} 114034

\bibitem{Inno} http$://$innolight.de$/$pdf$/$laser$\_$accessories.pdf

\bibitem{Naka} Nakajima K and Nakajima T 2004 {\it Proc.\ SPIE} {\bf 5567} 1385

\bibitem{Elli} Elliffe E J {\it et al} 2005 \CQG {\bf 22} 257

\bibitem{Dahl} Dahl K {\it et al} 2010 {\it J.\ Phys.:\ Conf.\ Ser.} {\bf 228} 012027

\bibitem{Dahl_bond} Dahl K {\it et al} Bonding accuracy of a suspension platform
interferometer with 23\,m long arms {\it in prep}

\bibitem{Hein} Heinzel G {\it et al} 2004 \CQG {\bf 21} S581

\bibitem{Shoe} Shoemaker D 2009 Advanced LIGO Reference Design {\it LIGO
Document} {\bf M060056-v1}\\ https$://$dcc.ligo.org$/$cgi-bin$/$DocDB$/$ShowDocument?docid=1507$\&$version=1

\bibitem{Morr} Morrison E, Meers B, Robertson D and Ward H 2004 {\it Appl.\ Opt.} {\bf33(22)} 5037--5041

\bibitem{opto} Schilling R 2010 OPTOCAD: (0.90c) A Fortran 95 module for tracing
Gaussian TEM$_{00}$ beams through an optical set-up \textit{Simulation tool}
\textit{http://www.rzg.mpg.de/$\sim$ros/optocad.html} 

\bibitem{Guzm} Guzm\'an Cervantes F, Steier F, Wanner G, Heinzel G and Danzmann K 2008 {\it Appl.\ Phys.\ B} {\bf90} 395



\end{thebibliography}
\end{document}